\documentclass[12pt]{article}

\usepackage{epsf}

\def\bfx{{\bf x}}
\def\bfy{{\bf y}}
\def\bfd{{\bf d}}
\def\bfi{{\bf i}}
\def\underd{{\underline{d}}}
\def\overmu{{\overline{\mu}}}
\def\overp{{\overline{p}}}
\def\OR{{\sc or}}
\def\AND{{\sc and}}
\def\NOT{{\sc not}}

\def\calD{{\cal D}}
\def\calA{{\cal A}}
\def\calB{{\cal B}}
\newtheorem{NoteAtEnd}{}

\title{Mathematical Basis for Physical Inference}
\author{Albert Tarantola\thanks{Institut de Physique du Globe;
                                4, place Jussieu; 
                                F-75005 Paris;
                                France; {\tt tarantola@ipgp.jussieu.fr}} 
     \ \& Klaus Mosegaard\thanks{Niels Bohr Institute for Astronomy, 
                                Physics and Geophysics; 
                                Dept.\ of Geophysics; 
                                Haraldsgade 6; 
                                DK-2200 Copenhagen N; 
                                Denmark; {\tt klaus@osiris.gfy.ku.dk}}}

\begin{document} 

\maketitle

% % % % % % % % % % % % % % % % % % % % % % % % % % % % % % % % % % % % % %
% Abstract
% % % % % % % % % % % % % % % % % % % % % % % % % % % % % % % % % % % % % %

\begin{abstract}
While the axiomatic introduction of a probability distribution over
a space is common, its use for making predictions, using physical 
theories and prior knowledge, suffers from a lack of formalization.
We propose to introduce, in the space of all probability distributions, 
two operations, the \OR\ and the \AND\ operation, that bring to the space 
the necessary structure for making inferences on possible values of
physical parameters.
While physical theories are often asumed to be analytical, we argue that
consistent inference needs to replace analytical theories by probability 
distributions over the parameter space, and we propose a systematic way of
obtaining such ``theoretical correlations'', using the \OR\ operation
on the results of physical experiments. 
Predicting the outcome of an experiment or solving ``inverse problems''
are then examples of the use of the \AND\ operation.
This leads to a simple and complete mathematical basis for general
physical inference.
\end{abstract}

% % % % % % % % % % % % % % % % % % % % % % % % % % % % % % % % % % % % % %
% Table of contents
% % % % % % % % % % % % % % % % % % % % % % % % % % % % % % % % % % % % % %

\clearpage
\vglue 40 mm
\tableofcontents
\clearpage

% % % % % % % % % % % % % % % % % % % % % % % % % % % % % % % % % % % % % %
% Section
% % % % % % % % % % % % % % % % % % % % % % % % % % % % % % % % % % % % % %

\section{Introduction}

Why has mathematical physics become so universal?
Does it allow the proper formulation of usual physical problems?
Some reasons explain the popularity of mathematical physics.
One reason is practical: mathematical physical theories 
may condensate a huge number of experimental results in a few 
functional relationships.
Perhaps more importantly, these relationships usually have a 
tremendous power of extrapolation, allowing the prediction of the
outcome of experiments never performed.
Psychologically, this capacity 
of predicting the outcome of experiments gives the very 
satisfactory feeling of ``understanding''.

Today, most scientists accept Popper's (\ref{note: Popper})
point of view that 
physics advances by postulating mathematical relations between
physical parameters. 
While it is fully recognized that physical theories should be 
confronted with experiments, Popper emphasized that sucessful 
predictions, in whatever number, can never prove that a theory is 
correct, but one single observation that contradicts the predictions 
of the theory is enough to refute, to falsify the whole theory. 
He also stressed that these contradictory results are of uttermost 
importance for the advance of physics. 
When Michelson and Morley (\ref{note: MichelsonAndMorley})
could not find the predicted difference in the speed of light
when the observer changes its velocity relative to the source, 
they broke the ground for the replacement of classical by 
relativistic mechanics.

%\eject

%\clearpage

Physical theories are conceptual models of reality. 
A good physical theory contains some or all of the following elements:

i) a modelisation of the space-time (for instance, as a four-dimensional 
continuum, or as a fractal entity)

ii) a modelisation of the objects of the ``universe'' (for instance, 
as point particles, or as continuous media)

iii) a recognition of the significant parameters in the experiments 
to be performed and a precise, operational, definition of these parameters

iv) mathematical relations postulated between these parameters, 
obtained by trying to obtain the best fit between observations and 
theoretical predictions.

While no physics is possible without points (i-iii) above, point (iv), 
i.e., postulating functional relations between the parameters to be 
used is not a necessity.

Any physical knowledge is uncertain, and estimation of uncertainties 
is crucial, for prosaic (e.g., preventing mechanical structures 
to collapse) as well as ethereal 
(e.g., using experimental results to decide between theories) reasons.
The problem we face is that while considerable effort is spent in
estimating experimental uncertainties, once a theory is postulated
that is acceptable in view of these uncertainties, usual mathematical
physics reasons as if the theory was exact. 
For instance, while we can use Gravitation Theory to predict the
behaviour of space-time near the big-bang of certain models of the 
Universe, we have no means of estimating how uncertain are our 
predictions.

This is more striking when using analytical theories to solve the
so-called ``inverse problems'' 
(\ref{note: Backus}--\ref{note: MosegaardAndTarantola}),
where data, a priori information, and ``physical theories'' have to be used
to make inferences about some parameters. The consideration of exact
theories leads, at best, to inaccurate estimations, at worst,
to mathematical inconsistencies (\ref{note: Unconsistency}).

This paper proposes an alternative to the common practice 
of postulating functional relationships between physical parameters.
In fact, we propose a mathematical formalization 
of pure empiricism, as opposed to mathematical rationalism.
Essentially, we suggest to replace functional relationships between 
physical parameters by well defined probability distributions over
the parameter space.

The proposed formalism will, in some sense, only be a sort of 
``tabulation'' of systematically performed experiments. 
In some aspects it will be less powerful
than the one obtained through the use of analytical theories
(it will not be able to extrapolate); 
in some aspects it will be more powerful
(it will be able to properly handle actual uncertainties).

We will show how experiments could be, 
at least in principle, systematically performed so 
that a probability distribution in the parameter space is obtained 
that contains, as an analytical theory, the observed correlations 
between physical parameters, but, in addition, contains the full 
description of the attached uncertainties.
We will also explain how, once such a theoretical probability 
distribution has been obtained, we can use it to make predictions 
---that will have attached uncertainties--- or to use data to solve 
general inverse problems.

To fulfill this project we need to complete classical probability theory.
Kolmogorov (\ref{note: Kolmogorov}) 
proposed an axiomatic introduction of the notion of
probability distribution over a space. The definition of conditional 
probability is then the starting point for doing inferences, as,
for instance, through the use of the Bayes theorem. 
But the space of all probability distributions (over a given 
space) lacks structure. We argue below that there are two natural 
operations, the \OR\ and the \AND\ operation, to be defined over the 
probability distributions, that create the necessary structure: that of 
an inference space.
We will see that while the \OR\ operation corresponds to an obvious
generalization of ``making histograms'' from observed results,
the \AND\ operation is just the right generalization of the notion of
conditional probability.

%\clearpage

% % % % % % % % % % % % % % % % % % % % % % % % % % % % % % % % % % % % % %
% Section
% % % % % % % % % % % % % % % % % % % % % % % % % % % % % % % % % % % % % %

\section{The structure of an Inference Space}

Before Kolmogorov (\ref{note: Kolmogorov}), 
probability calculus was made using the 
intuitive notions of ``chance'' or ``hazard''. Kolmogorov's axioms
clarified the underlying mathematical structure and brought
probability calculus inside well defined mathematics.
In this section we will recall these axioms.
Our opinion is that the use in physical theories 
(where we have invariance requirements) of probability distributions, 
through the notions of conditional probability or the so-called 
Bayesian paradigm suffers today from the same defects 
as probability calculus suffered from before Kolmogorov.
To remedy this, we introduce in this section, in the space of all 
probability distributions, two logical operations (\OR\ and \AND) 
that give the necessary mathematical structure to the space.

%\clearpage

% % % % % % % % % % % % % % % % % % % % % % % % % % % % % % % % % % % % % %
% Section
% % % % % % % % % % % % % % % % % % % % % % % % % % % % % % % % % % % % % %

\subsection{Kolmogorov's concept of probability}

\label{sec: Kolmo}

A point \ $\bfx$~, that can
materialize itself anywhere inside a domain \ $\calD$~, may be realized,
for instance, inside \ $ \calA $~, a subdomain of \ $ \calD $~.
The probability of realization of the point is completely described if we
have introduced a {\em probability distribution\/} 
(in Kolmogorov's [\ref{note: Kolmogorov}] sense) 
on \ $ \calD $~, i.e., if to every subdomain \ $ \calA $ \ 
of \ $\calD$ \ we are able to associate a real number \ $P(\calA) $~,
called {\em the probability of\/} \ $ \calA $~, having the three
properties:

\medskip

\noindent\ $ \bullet $ \ 
\hskip -3 pt 
For any subdomain \ $ \calA $ \ of \ $ \calD$~, \ $P(\calA)\geq 0$~.

\noindent\ $ \bullet $ \ 
\hskip -3 pt 
If \ $\calA_i$ \ and \ $\calA_j$ \ are two disjoint subsets of 
\ $ \calD $~, then, 
\ $ P(\calA_i \cup \calA_j) = P(\calA_i) + P(\calA_j) $~.

\noindent\ $ \bullet $ \ 
For a sequence of events 
\ $\calA_1 \supseteq \calA_2 \supseteq \cdots $ \ 
tending to the empty set, 
we have \ $ P(\calA_i)\rightarrow 0 $~.

\medskip

We will not necessarily assume that a probability distribution 
is normed to unity (~$ P(\calD)=1 $~). Although one refers to this
as a {\em measure\/}, instead of a probability, we will not use this 
distinction. Sometimes, our probability distributions will not be 
normalizable at all (~$ P(\calD) = \infty $~). We can only then compute
the {\em relative probabilities\/} of subdomains.

These axioms apply to probability distributions over discrete or
continuous spaces. Below, we will consider probability distributions
over spaces of physical parameters, that are continuous spaces.
Then, a probability distribution is represented by a probability density
(note [\ref{note: Radon-Nicodym}] explains the difference between
a probability density and a volumetric probability).

In the next section, given a space \ $ \calD $~, we will consider 
different probability distributions \ $ P \, , \, Q \dots$ \ 
Each probability distribution will represent a particular 
{\em state of information\/} over \ $ \calD $~.
In what follows, we will use as synonymous the terms 
``probability distribution'' and ``state of information''.

%\clearpage

% % % % % % % % % % % % % % % % % % % % % % % % % % % % % % % % % % % % % %
% Section
% % % % % % % % % % % % % % % % % % % % % % % % % % % % % % % % % % % % % %

\subsection{Inference space}

We will now give a structure to the space of all the probability 
distributions over a given space, by introducing two operations, 
the \OR\ and the \AND\ operation. This contrasts with the basic
operations introduced in deductive logic, where the negation
(``\NOT''), nonexistent here, plays a central role.
In what follows, the \OR\ and the \AND\ operation will be denoted,
symbollically, by \ $\vee $ \ and \ $\wedge $~. 
They are assumed to satisfy the set of axioms here below.

The first axiom states that if an event \ $ \calA $ \ is possible for
\ $ (P $\/ \OR\ $ Q) $~, then the event is either possible for
\ $ P $ \ or possible for \ $ Q $ \ 
(which his is consistent with the usual logical sense for the ``or''):
{\em For any subset \ $\calA$~, and for any two probability
distributions \ $ P $ \ and \ $ Q $~, the \OR\ operation satisfies
$$
\left( P\vee Q\right) (\calA) \neq 0 \qquad \Longrightarrow \qquad
 P(\calA)\neq 0 \quad {\rm or} \quad  Q(\calA) \neq 0 \ ,
$$
the word {\rm ``or''} having here its ordinary logical sense.\/}

The second axiom states that if an event \ $ \calA $ \ is possible for
\ $ (P $\/ \AND\ $ Q) $~, then the event is possible for
both \ $ P $ \ and \ $ Q $ \ 
(which is consistent with the usual logical sense for the ``and''): 
{\em For any subset \ $\calA$~, and for any two probability
distributions \ $ P $ \ and \ $ Q $~, the \AND\ operation satisfies
$$
\left( P \wedge Q \right) (\calA) \neq 0 \qquad \Longrightarrow \qquad 
P(\calA) \neq 0 \quad {\rm and} \quad  Q(\calA) \neq 0 \ ,
$$
the word {\rm ``and''} having here its ordinary logical sense.\/}

The third axiom ensures the existence of a neutral element, that
will be interpreted below as the probability distribution carrying
no information at all:
{\em There is a {\em neutral element\/}, \ $M$ \ for the \AND\ 
operation, i.e., it exists a \ $M$ \ such that for any probability
distribution \ $P$ \ and for any subset \ $\calA$~,\/}
$$
\left( M\wedge P\right) (\calA) = \left( P\wedge M\right) (\calA)
= P(\calA) \ .
$$

The fourth axiom imposes that {\em the \OR\ and the \AND\  operations 
are {\em commutative\/} and {\em associative\/}\/},
and, by analogy with the algebra of propositions of ordinary logic, 
have a distributivity property: {\em the \AND\ operation is 
{\em distributive\/} with respect to the
\OR\ operation\/}.

The structure obtained when furnishing 
the space of all probability distributions 
(over a given space \ $ \calD $~)
with two operations \OR\ and \AND, 
satisfying the given axioms constitutes what we propose to call 
an {\em inference space\/}.

These axioms do not define uniquely the operations.
Let \ $ \mu(\bfx) $ \ be the particular probability density representing
\ $ M $~, the neutral element for the \AND\ operation, and let
\ $ p(\bfx) , q(\bfx)\dots $ \ be the probability densities representing
the probability distributions \ $ P, Q\dots $ \ 
Using the notations \ $ (p \vee q)(\bfx) $ \ and
\ $ (p \wedge q)(\bfx) $ \ for the probability densities
representing the probability distributions
\ $ P \vee Q $ \ and \ $ P \wedge Q $ \ respectively,
one realization of the axioms (the one we will retain) is given by
\begin{equation}
(p \vee q)(\bfx) = p(\bfx) + q(\bfx)
\qquad ; \qquad
(p \wedge q)(\bfx) = \frac{p(\bfx) \, q(\bfx)}{\mu(\bfx)} \ ,
\label{eq: WeWant}
\end{equation}
where one should remember that we do not impose to our probability 
distributions to be normalized.

The structure of an inference space, as defined, contains other
useful solutions. 
For instance, the theory of fuzzy sets (\ref{note: Kandel})
uses positive functions \ $ p(\bfx) , q(\bfx) \dots $ \ 
quite similar to probability densities, but having a different
interpretation: the are normed by the condition that their maximum value
equals one, and are interpreted as the ``grades of membership'' 
of a point \ $ \bfx $ \ to the ``fuzzy sets'' \ $ P , Q \dots $~.
The operations \OR \ and \AND \ correspond then respectively 
to the {\em union\/} and {\em intersection\/} of fuzzy sets, 
and to the following realization of our axioms:
\begin{equation}
(p \vee q)(\bfx) = {\rm max}\!\left(p(\bfx) , q(\bfx)\right)
\qquad ; \qquad
(p \wedge q)(\bfx) = {\rm min}\!\left(p(\bfx) , q(\bfx)\right) \ ,
\label{eq: WeNotWant}
\end{equation}
where the neutral element for the \AND \ operation 
(intersection of fuzzy sets) is simply the function
\ $ \mu(\bfx) = 1 $~.

While fuzzy set theory is an alternative to classical probability
(and is aimed at the solution of a different class of problems), 
our aim here is only to complete the classical probability theory. 
As explained below the solution given by equations~\ref{eq: WeWant}
correspond to the natural generalisation of two fundamental operations
in classical probability theory: that of ``making histograms'' and
that of taking ``conditional probabilities''.
To simplify our language, we will sometimes use this correspondence 
between our theory and the fuzzy set theory,
and will say that the \OR\ operation, when applied to two 
probability distributions, corresponds to the {\em union\/} 
of the two states of information, while the \AND\ operation 
corresponds to their {\em intersection\/}.

It is easy to write some extra conditions that distinguish 
the two solutions given by
equations~\ref{eq: WeWant} and~\ref{eq: WeNotWant}.
For instance, as probability densities are normed using a 
multiplicative constant (this is not the case with the grades of 
membership in fuzzy set theory), it makes sense to impose the simplest
possible algebra for the multiplication of probability densities 
\ $ p(\bfx) , q(\bfx) \dots $ \ 
by constants \ $ \lambda , \mu \dots $~:
\begin{equation}
\left[ (\lambda+\mu) p \right] (\bfx) 
= \left( \lambda \, p \vee \mu p \right) (\bfx)
\qquad ; \qquad
\left[ \lambda ( p \wedge q) \right] (\bfx) 
= 
\left( \lambda p \wedge q \right) (\bfx) 
= 
\left( p \wedge  \lambda q \right) (\bfx) \ .
\end{equation}
This is different from finding a (minimal) set of axioms characterizing 
(uniquely) the proposed solution, which is an open problem.

One important property of the two operations \OR \ and \AND \ 
just introduced is that of {\em invariance\/} with respect
to a change of variables.
As we consider probability distribution over a continuous space,
and as our definitons are independent of any choice of coordinates
over the space, it must happen that we obtain equivalent results 
in any coordinate system. Changing for instance from the coordinates 
\ $ \bfx $ \ to some other coordinates \ $ \bfy $~,
will change a probability density \ $ p(\bfx) $ \ to 
\ $ \widetilde{p}(\bfy) = p(\bfx) \, |\partial\bfx/\partial\bfy | $~.
It can easily be seen (\ref{note: DemoInvariance})
that performing the \OR\ or the \AND\ operation,
then changing variables, gives the same result than first changing variables,
then, performing the \OR\ or the \AND\ operation.

Let us mention that the equivalent of equations~\ref{eq: WeWant}
for discrete probability distributions is:
\begin{equation}
\left( p\vee q\right)_i = p_i + q_i 
\qquad ; \qquad 
\left( p \wedge q \right)_i = \frac{p_i \, q_i}{\mu_i} \ .
\end{equation}

Although the \OR\ and \AND\ notions just introduced
are consistent with classical logic, they are here more general, 
as they can handle states of information that are more subtle
than just the ``possible'' or ``impossible'' ones.

%\clearpage

% % % % % % % % % % % % % % % % % % % % % % % % % % % % % % % % % % % % % %
% Section
% % % % % % % % % % % % % % % % % % % % % % % % % % % % % % % % % % % % % %

\subsection{The interpretation of the \OR\ and the \AND\ operation}

If an experimenter faces realizations of a random process and wants
to investigate the probability distribution governing the process,
he may start making histograms of the realizations. For instance,
for realizations of a probability distribution over a continuous space,
he will obtain histograms that, in some sense, will approach the 
probability density corresponding to the probability distribution.

A histogram is typically made by dividing the working space into cells, 
and by counting how many realizations fall inside each cell.
A more subtle approach is possible. First, we have to understand that,
in the physical sciences, when we say ``a random point
has materialized in an abstract space'', we may mean something like 
``this object, one among many that may exist, 
vibrates with some fixed period; let us measure as accurately as possible
its period of oscillation''.
Any physical measure of a real quantity will have attached
uncertainties. As explained in section~\ref{sec: Measuring},
this means that when, mathematically speaking,
we measure ``the coordinates of a point in an abstract space''
we will not obtain a point, but a state of information over the space,
i.e., a probability distribution.

\begin{figure}[htbp]
\epsfxsize 10 cm
\centerline{\epsffile{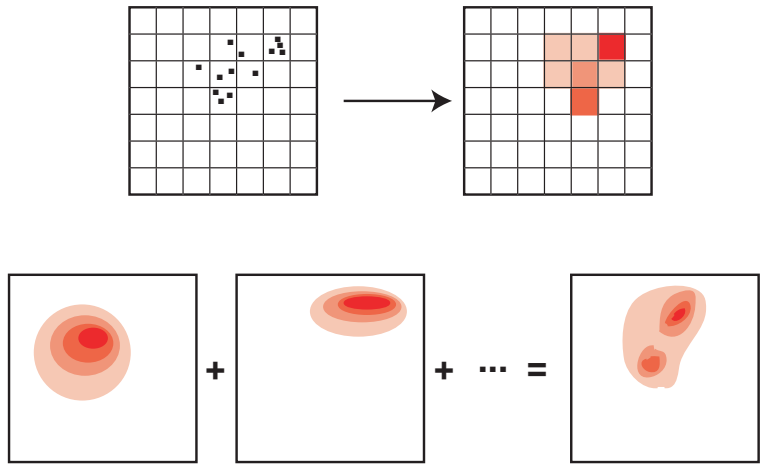}}
\caption{Illustration of the \OR\ operation applied to probability 
         distributions.
         A histogram is made (see top of the figure) by dividing the working 
         space into cells,
         and by counting how many realizations fall inside each cell.
         A more subtle approach is possible.
         First, we have to understand that, in the physical sciences, 
         when we say ``a random point has materialized in an abstract 
         space'', we may mean something like ``this object, one among 
         many that may exist, vibrates with some fixed period; 
         let us measure as accurately as possible its period of 
         oscillation''. Any physical measure of a real quantity will 
         have attached uncertainties. This means that when, mathematically 
         speaking, we measure ``the coordinates of a point in an abstract 
         space'' we will not obtain a point, but a state of information 
         over the space, i.e., a probability distribution. 
         If we have measured the coordinates of many points, 
         the results of each measurement will be described by a 
         probability density \ $ p_i(x) $~. The union of all these, 
         i.e., the probability density 
         \ $ (p_1 \vee p_2 \vee \dots )(\bfx) = \sum_i p_i(\bfx) $ \ 
         is a finer estimation of the background probability density 
         than an ordinary histogram, as actual measurement uncertainties 
         are used, irrespectively of any division of the space into cells. 
         If it happens that the measurement uncertainties can be 
         described using box-car functions (at fixed positions), 
         then, the approach we propose reduces to the conventional 
         making of histograms.}
\label{fig: Histogram}
\end{figure}

If we have measured the coordinates of many points, 
the results of each measurement will be described by a probability 
density
\ $ p_i(\bfx) $~. 
The union of all these, i.e., the probability density
\begin{equation}
\left(p_1 \vee p_2 \vee \dots \right) (\bfx) = \sum_i p_i(\bfx)
\end{equation}
is a finer estimation of the background probability density
than an ordinary histogram, 
as actual measurement uncertainties are used, irrespectively 
of any division of the space into cells. If it happens that 
the measurement uncertainties can be described using box-car 
functions at fixed positions, then, the approach we propose 
reduces to the conventional making of histograms.
This is illustrated in figure~\ref{fig: Histogram}.

% % % % % % % % % % % % % % % % % % % % % % % % % % % % % % % % % % % % % %
% Begin Figure
% % % % % % % % % % % % % % % % % % % % % % % % % % % % % % % % % % % % % %

% % % % % % % % % % % % % % % % % % % % % % % % % % % % % % % % % % % % % %
% Figure
% % % % % % % % % % % % % % % % % % % % % % % % % % % % % % % % % % % % % %

\begin{figure}[htbp]
\epsfxsize \textwidth
\centerline{\epsffile{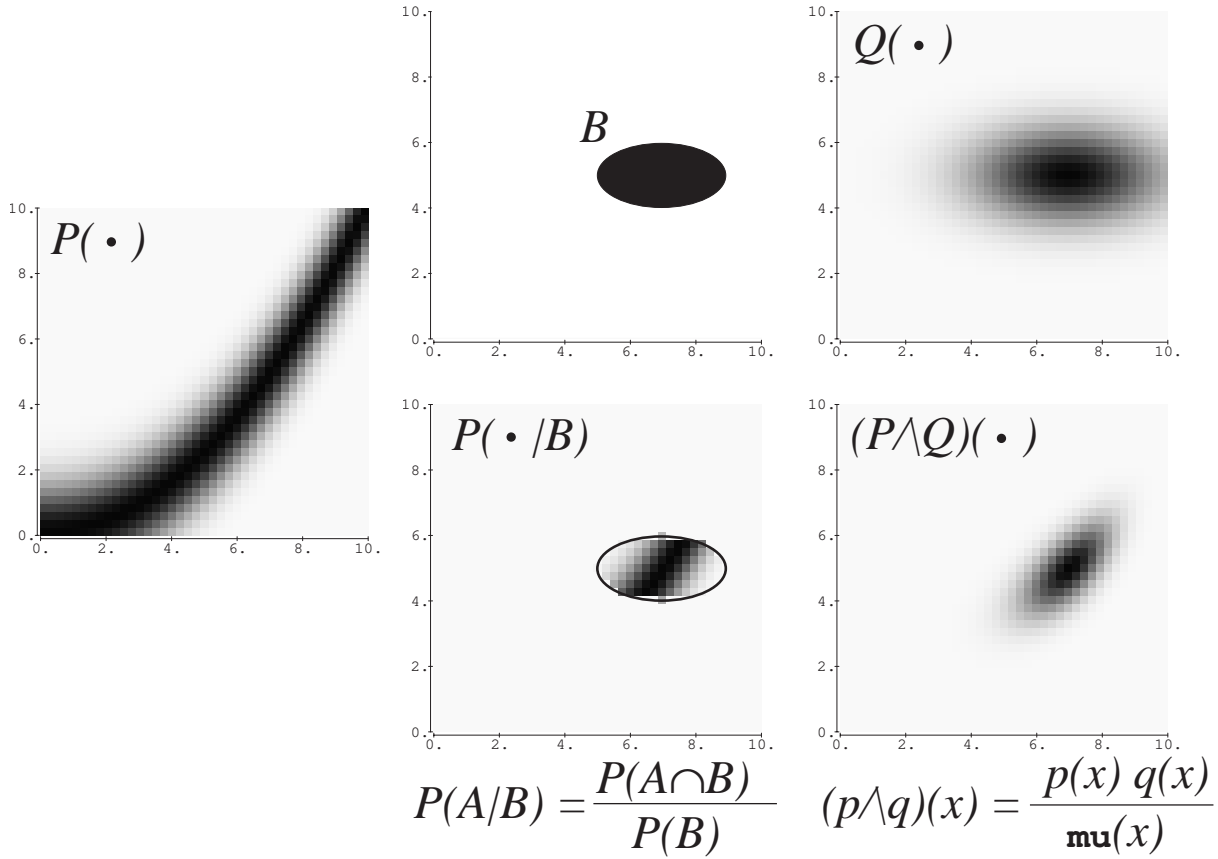}}
\caption{Illustration of the \AND\ operation applied to probability 
         distributions.
         This figure explains that our definition of the \AND\ 
         operation is a generalization of the notion of conditional 
         probability.
         A probability distribution \ $ P(\ \cdot\ ) $ \ 
         is represented (left of the figure) by its probability density. 
         To any region \ $ \calA $ \
         of the plane, it associates the probability \ $ P(\calA) $~.
         If a point has been realized following the probability 
         distribution \ $ P(\ \cdot\ ) $ \ and we are given the 
         information that, in fact, the point 
         is ``somewhere'' inside the region \ $ \calB $~, 
         then we can update the prior probability \ $ P(\ \cdot\ ) $~, 
         replacing it by the conditional
         probability \ $ P(\ \cdot\ |\calB) = 
         P(\ \cdot\ \cap\calB)/P(\calB) $~.
         It equals \ $ P(\ \cdot\ ) $ \ inside \ $ \calB $ \ 
         and is zero outside (center of the figure).
         If instead of the hard constraint \ $ x \in \calB $ \ 
         we have a soft information about the location of \ $ x $~, 
         represented by the probability distribution \ $ Q(\ \cdot\ ) $ \ 
         (right of the figure), the intersection of the two states of 
         information \ $ P $ \ and \ $ Q $ \ 
         gives a new state of information (here, \ $ \mu(x) $ \
         is the probability density representing the state 
         of null information, and, to simplify the figure, 
         has been assumed to be constant).
         The comparison of the right with the center of the figure
         shows that the \AND\ operation generalizes the notion of 
         conditional probability.
         In the special case where the probability density representing
         the second state of information, \ $ Q(\ \cdot\ ) $~, 
         equals the null information probability density inside 
         the domain \ $ \calB $ \ and is zero outside, then, the notion 
         of intersection of states of information exactly reduces to 
         the notion of conditional probability.}
\label{fig: ConDens}
\end{figure}

% % % % % % % % % % % % % % % % % % % % % % % % % % % % % % % % % % % % % %
% End Figure
% % % % % % % % % % % % % % % % % % % % % % % % % % % % % % % % % % % % % %

Figure~\ref{fig: ConDens} explains that our definition of the \AND\ 
operation is a generalization of the notion of conditional probability.
A probability distribution \ $ P(\ \cdot\ ) $ \ 
is represented, in the figure, by its probability density. 
To any region \ $ \calA $ \
of the plane, it associates the probability \ $ P(\calA) $~.
If a point has been realized following the probability distribution
\ $ P(\ \cdot\ ) $ \ and we are given the information that, in fact, 
the point 
is ``somewhere'' inside the region \ $ \calB $~, then we can update
the prior probability \ $ P(\ \cdot\ ) $~, replacing it by the conditional
probability \ $ P(\ \cdot\ |\calB) = P(\ \cdot\ \cap\calB)/P(\calB) $~.
It equals \ $ P(\ \cdot\ ) $ \ inside \ $ \calB $ \ and is zero outside
(center of the figure).
If instead of the hard constraint \ $ x \in \calB $ \ 
we have a soft information about the location of \ $ x $~, 
represented by the probability distribution \ $ Q(\ \cdot\ ) $ \ 
(right of the figure), the intersection of the two states of 
information \ $ P $ \ and \ $ Q $ \ gives a new state of information
(here, \ $ \mu(x) $ \
is the probability density representing the state of null information,
and, to simplify the figure, has been assumed to be constant).
The comparison of the right with the center of the figure
shows that the \AND\ operation generalizes the notion of conditional 
probability.
In the special case where the probability density representing
the second state of information, \ $ Q(\ \cdot\ ) $~, equals the null
information probability density inside the domain \ $ \calB $ \ 
and is zero outside, then, the notion of intersection of states
of information exactly reduces to the notion of conditional probability.

Now the interpretation of the neutral element for the \AND\ 
operation can be made clear. We postulated that the neutral 
probability distribution \ $ M $ \ is such that for any probability 
distribution \ $ P $~, \ $ P \wedge M = P $~. This means that
if a point is realized according to a probability distribution \ $ P $~,
and if a (finite accuracy) measure of the coordinates
of the point produces the information represented by \ $ M $~, 
the posterior probability distribution, \ $ P \wedge M $ \ 
is still \ $ P $~: the probability distribution \ $ M $ \ 
is not carrying any information at all.
Accordingly, we call \ $ M $ \ the {\em null information probability
distribution\/}.
Sometimes, the probability density representing this state
of null information is constant over all the space; 
sometimes, it is not, as explained in section~\ref{sec: NonInf}.
It is worth mentioning that this particular
state of information enters in the Shannon's definition of 
Information Content (\ref{note: Shannon-1},~\ref{note: Shannon-2}).

It is unfortunate that, when dealing with probability distributions over
continous spaces, conditional probabilities are often misused. 
Note (\ref{note: CondProb}) describes the so-called Borel-Kolmogorov
paradox: using conditional probability densities in a space with
coordinates \ $ (x,y) $ \ will give results that will not be consistent
with those obtained by the use of conditional probability densities
on the same space but where other coordinates \ $ (u,v) $ \ are used
(if the change of coordinates is nonlinear).
Jaynes (\ref{note: Jaynes}) gives an excellent, explicit,
account of the paradox. But his choice for resolving the paradox is 
different from our's: while Jaynes just insists on the technical
details of how some limits have to be taken in order to ensure
consistency, we radically decide to abandon the notion of conditional
probability, and replace it by the intersection of states of information
(the \AND\ operation) which is naturally consistent under a change of 
variables, as demonstrated in note (\ref{note: DemoInvariance}).

%\clearpage

% % % % % % % % % % % % % % % % % % % % % % % % % % % % % % % % % % % % % %
% Section
% % % % % % % % % % % % % % % % % % % % % % % % % % % % % % % % % % % % % %

\section{Physical parameters}

Crudely speaking, a physical parameter is anything that can be measured.
For a physical parameter, like a temperature, an electric field, 
or a mass, can only be defined by prescribing the experimental 
procedure that will measure it. Cook (\ref{note: Cook}) discusses this 
point with lucidity.

The theory to be developed in this article will be illustrated by 
the analysis of objects that have a characteristic length, \ $ L $~, 
affected by phenomena that have a characteristic period, \ $ T $~.
A measurement of a parameter is performed by
realizing the conventional unit (i.e., the meter for a length, 
the second for a duration) and by comparing the parameter to the unit.
We have then to turn to the definition of the units of time duration 
and of length.

At present, the {\em second\/} is defined
as the duration of \ 9 192 631 770 periods of the radiation 
corresponding to the transition between the two hyperfine 
levels of the ground state of the c{\ae}sium-133 atom. 
Practically this means that a beam of c{\ae}sium-133 atoms are 
submitted to an electromagnetic field of adjustable frequency: 
when the imposed frequency is such that it causes the transition 
between the two hyperfine levels of the ground state of the atoms, 
the standard of frequency (and, thus, of period) has been realised.

Until 1991, the unit of length used to be defined independently of 
that of time duration. Now the {\em meter\/} is connected 
to the second by defining the value of the velocity of light as
\ $ c = $ 299 792 458 m\,s$^{-1}$. 
This means, in fact, that lengths are measured by measuring the 
time it takes light to traverse them (and then, converting to 
distance through this conventional value of \ $ c $~).

%\clearpage 

% % % % % % % % % % % % % % % % % % % % % % % % % % % % % % % % % % % % % %
% Section
% % % % % % % % % % % % % % % % % % % % % % % % % % % % % % % % % % % % % %

\subsection{The noninformative probability density for physical parameters}
\label{sec: NonInf}

Once a physical parameter has been defined, it is possible to 
associate to it a particular probability distribution, that will 
represent, when making a measurement, the absence of information 
on the possible outcome of the experiment.

Assume that, furnished with our definition of the unit of time 
duration, we wish to measure the period of some object. 
It can be the period of a rotating galaxy, or the period of a 
XVII-th century pendulum, or the period of a vibrating molecule: 
we do not know yet. Let us denote by \ $ p(T) $ \ 
the probability density representing this state of total ignorance. 
The frequency \ $ \nu $ \ associated to the period \ $ T $ \ is
\ $ \nu = 1/T $~.
>From \ $ p(T) $ \ we can, using the general rule of change of variables, 
deduce the probability density for the frequency:
\ $ q(\nu) 
= p(T) \, \left| dT/d\nu \right|
= p(T) / \nu^2  $~. 

Now, the definition of the unit of time duration 
is undistinguishable from the definition of the unit of frequency.
In fact, when trying to define the standard of time we said 
``when the imposed frequency is such that is causes the transition 
between the two hyperfine levels of the ground state of the atoms, 
the standard of {\em frequency\/} has been realised'', which shows 
how closely related are the reciprocal parameters period-frequency: 
we can not define the unit {\em second\/} without defining, 
at the same time, the unit {\em Hertz\/}.

We find here, at a very fundamental level, the class of reciprocal 
parameters analyzed by Harold Jeffreys (\ref{note: Jeffreys}). 
As he argued, the null information probability density must have 
the same form for the two parameters, i.e., 
\ $ p(\cdot) $ \ and \ $ q(\cdot) $ \ must be 
{\em the same function\/}. Then, 
the constraint
\ $ q(\nu) = p(T) / \nu^2  $~,
seen above, gives, 
up to a multiplicative constant, the solution 
\begin{equation}
p(T) = \frac{1}{T} 
\qquad ; \qquad 
q(\nu) = \frac{1}{\nu}
\ . 
\label{eq: TheSol}
\end{equation}

The range of time durations (or of periods) 
considered in physics spans many orders of magnitude 
(from periods of atomic objects to cosmological periods).
Physicists then often use a logarithmic scale, defining,
for instance,
\ $ T^\ast = \log(T/T_0) $ \ 
and
\ $ \nu^\ast = \log(\nu/\nu_0) $~,
where the two constants \ $ \nu_0 $ \ and \ $ T_0 $ \ can be 
arbitrary (\ref{note: TheConstants}).
Transforming the probability densities in~\ref{eq: TheSol} 
to the logarithmic variables gives
\ $ p^\ast(T^\ast) = 1 $ \ 
and
\ $ q^\ast(\nu^\ast) = 1 $~.
The logarithmic variables (that take values on all the real line) 
have a constant probability density. 
This, is fact, is the deep interpretation of the \ $ 1/x $ \ 
probability densities in equations~\ref{eq: TheSol}.
The particular variables for which the probability density
representing the state of null information is a constant over
all the space can be named {\em Cartesian\/}: they are
more ``natural'' than others, as are the usual Cartesian coordinates 
in Euclidean spaces~(\ref{note: Cartesian}).
That these ``Cartesian'' variables are not only 
more natural, but also more practical than other variables, 
can be understood by considering 
that manufacturers of pianos space notes with constant increments not of 
frequency, but of the associated logarithmic variable.

We have seen that the definition of length is today related to that 
of time duration through the velocity of light. We could say that
the electromagnetic wave of the radiation that defines the unit of time,
defines, through its wavelength, the unit of distance. But, here again,
we have a perfect symmetry between the wavelength and its inverse, 
the wavenumber. This is why we take the function \ $ h(L) = 1/L $ \ 
to describe the null information probability density for the length
of an object (\ref{note: amusing}).

%\clearpage

% % % % % % % % % % % % % % % % % % % % % % % % % % % % % % % % % % % % % %
% Section
% % % % % % % % % % % % % % % % % % % % % % % % % % % % % % % % % % % % % %

\subsection{Measuring physical parameters}
\label{sec: Measuring}

To define the experimental procedure that will lead to a 
``measurement'' we need to
conceptualize the objects of the ``universe'': do we have
point particles or a continuous medium?
Any instrument that we can build will have
finite accuracy, as any manufacture is imperfect.
Also, during the measurement act,
the instrument will always be submitted to unwanted sollicitations 
(like uncontrolled vibrations).

This is why, even if the experimenter postulates the existence of
a well defined, ``true value'', of the measured parameter, she/he
will never be able to measure it exactly.
Careful modeling of experimental uncertainties is not easy,
Sometimes, the result of a measurement of a parameter \ $ p $ \ 
is presented as 
\ $ p = p_0 \pm \sigma $~, 
where the interpretation of \ $ \sigma $ \ may be diverse. 
For instance, the experimenter may imagine a bell-shaped
probability density around \ $ p_0 $ \ representing her/his
state of information ``on the true value of the parameter''.
The constant \ $ \sigma $ \ can be the standard deviation 
(or mean deviation, or other estimator of dispersion) of
the probability density used to model the experimental
uncertainty.

In part, the shape of this probability density may come from
histograms of observed or expected fluctuations. In part,
it will come from a subjective estimation of
the defects of the unique pieces of the instrument.
We postulate here that the result of any measurement can,
in all generality, be described by defining a probability
density over the measured parameter, representing the 
information brought by the experiment on the ``true'', 
unknowable, value of the parameter.
The official guidelines for expressing uncertainty in measurement,
as given by the International Organization for Standardization (ISO)
and the National Institute of Standards and Technology (\ref{note: ISO})
although stressing the special notion of standard deviation, 
are consistent with the possible use of general probability 
distributions to express the result of a measurement, as advocated here.

Any shape of the density function is not acceptable.
For instance, the use of a Gaussian density to
represent the result of a measurement of a positive 
quantity (like an electric resistivity) would give
a finite probability for negative values of the variable,
which is inconsistent (a lognormal probability density,
on the contrary, could be acceptable).

In the event of an ``infinitely bad measurement'' 
(like when, for instance, an unexpected
event prevents, in fact, any meaningful measure) the result
of the measurement should be described using the null information
probability density introduced above. 
In fact, when the density function used to represent the result
of a mesurement has a parameter \ $ \sigma $ \ describing the
``width'' of the function, it is the limit of the density
function for \ $ \sigma \to \infty $ \ that should represent
a measurement of infinitely bad quality.
This is consistent, for instance, with the use of a lognormal
probability density for a parameter like an electric 
resisitivity \ $ r $~, as the limit of the lognormal for 
\ $ \sigma \to \infty $ \ is the \ $ 1/r $ \ function,
which is the right choice of noninformative probability density
for~$ r $~.

Another example of possible probability density to represent
the result of a measurement of a parameter \ $ p $ \ 
is to take the noninformative probability density for
\ $ p_1 < p < p_2 $ \ and zero outside. 
This fixes strict bounds for possible values of the parameter, 
and tends to the noninformative probability density 
when the bounds tend to infinity.

The point of view proposed here will be consistent with the
the use of ``theoretical parameter correlations''
as proposed in section \ref{sec: UsingBayesTheory}, so that there
is no difference, from our point of view, between a 
``simple measurement'' and a measurement using physical theories, 
including, perhaps, sophisticated inverse methods.

%\clearpage

% % % % % % % % % % % % % % % % % % % % % % % % % % % % % % % % % % % % % %
% Section
% % % % % % % % % % % % % % % % % % % % % % % % % % % % % % % % % % % % % %

\section{Bayesian physical theories}

Physical ``laws'' prevent us from setting arbitrarily some 
physical parameters. For instance, we can set the length of a 
tube where a free fall experiment will be performed, and we can 
also decide on the place and time of the experiment, but the 
time duration of the free fall is ``imposed by Nature''.
Physics in much about the analysis of these physical correlations
between parameters.

Typically, a set \ $ \bfi $ \ of {\em independent parameters\/} 
is identified, and experiments are performed in order to measure 
the values of a set \ $ \bfd $ \ of {\em dependent parameters\/}
(\ref{note: FreeWill}).
Analytical physical theories try then to express the result of
the observations by a functional relationship \ $ \bfd = \bfd(\bfi) $~.
In fact, saying that the independent parameters are ``set''
and the dependent parameters ``measured'' is an oversimplification,
as all the parameters must be measured.
And, as discussed in the previous section, 
uncertainties are present in every measurement. 
The values of the parameters that are set (the independent parameters) 
are never known exactly. The measures of the dependent parameters 
have always uncertainties attached. Assume we have made a large
number of experiments, that show how the dependent parameters 
correlate with the independent ones.
Within the error bars of the experimental results it will always be
possible to fit an infinity of functional relationships of the
form \ $ \bfd = \bfd(\bfi) $~.
Adding more experimental points may 
help to discard some of the ``theories'', but there will always remain
an infinity of them. 

We formalize this fact at a fundamental level, 
by replacing the need of a functional relationship by the 
use of a probability distribution in the space of all the 
parameters considered, representing the actual information 
we may have.
Not only this point of view corresponds to a certain philosophy 
of physics, it also leads ---as discussed below--- 
to the only consistent formalism we know that is able to 
predict values of possible observations and of the attached 
uncertainties.

To be complete, we consider two cases where we may wish to analyze 
the physical correlations between parameters. The first case is
when a repetitive phenomenon takes place spontaneously. The second
case correspond to the case when an experimenter prompts a physical
phenomenon, using an experimental arrangement.

%\clearpage

% % % % % % % % % % % % % % % % % % % % % % % % % % % % % % % % % % % % % %
% Section
% % % % % % % % % % % % % % % % % % % % % % % % % % % % % % % % % % % % % %

\subsection{The ``contemplative'' point of view}

Consider an astronomer trying to analyze the ``relationship''
between the initial magnitude \ $ m $ \ of shooting stars and 
the total distance \ $ \Delta $ \ traveled by the meteors 
on the sky before disintegration.
Each shooting star naturally appearing on the sky will allow
one measurement of the two parameters \ $ m $ \ and \ $ \Delta $ \ 
to be performed (and possibly other significant parameters). 
As discussed above, each result of a measurement
will be represented by a probability density. 
Let \ $ \theta_i(m,\Delta) $ \ be the probability density representing
the information obtained on the parameters \ $ m $ \ and \ $ \Delta $ \ 
of the \ $i$-th \ shooting star.

When a large enough number of shooting starts has been observed, the
correlation between the parameters  \ $ m $ \ and \ $ \Delta $ \  is
perfectly described by the probability density obtained by 
applying the \OR\ operation 
(as defined by the first of equations~\ref{eq: WeWant})
to the probability distributions 
represented by \ $ \theta_1(m,\Delta) , \theta_2(m,\Delta) , \dots $~,
i.e., by the probability density
\ $ \theta(m,\Delta) = \sum_i \theta_i(m,\Delta) $~.
If, more generally, the observed parameters are generically represented 
by \ $ \bfx $~, and the result of the \ $i$-th \ experiment, 
by the probability density \ $ \theta_i(\bfx) $~, then,
\begin{equation}
\theta(\bfx) = \sum_i \theta_i(\bfx) \ .
\label{eq: GoingToTheta}
\end{equation}
The utility of this probability density will be explained
in section~\ref{sec: UsingBayesTheory}.

%\clearpage

% % % % % % % % % % % % % % % % % % % % % % % % % % % % % % % % % % % % % %
% Section
% % % % % % % % % % % % % % % % % % % % % % % % % % % % % % % % % % % % % %

\subsection{The ``experimental'' point of view}

Here, the independent parameters \ $ \bfi $ \ are ``set'',
and the dependent parameters \ $ \bfd $ \ measured.
This case can be reduced to the previous case (the ``contemplative''
one) provided that the independent parameters \ $ \bf i $ \ 
are ``randomly generated'' according to some reference probability
distribution, as, for instance, the null information probability
distribution discussed in section~\ref{sec: NonInf}
(this guaranteeing, in particular, that any possible region of 
the space of independent parameters will eventually be sampled).

As above, if \ $ \theta_i(\bfi,\bfd) $ \ is the probability density 
representing the information on \ $ \bfi $ \ and \ $ \bfd $ \ 
obtained from the \ $i$-th \ experiment, after a large enough
number of experiments has been performed, the correlations
between the dependent and the independent parameters are
described by the probability density 
\ $ \theta(\bfi,\bfd) = \sum_i \theta_i(\bfi,\bfd) $~.
In general, if the whole set of parameters is generically 
represented by \ $ \bfx = \{\bfi,\bfd\} $~, and the result of 
the \ $i$-th \ experiment, by the probability density 
\ $ \theta_i(\bfx) $~, then equation~\ref{eq: GoingToTheta} 
holds again.

\begin{figure}[htbp]
\epsfxsize 65 mm
\centerline{\epsffile{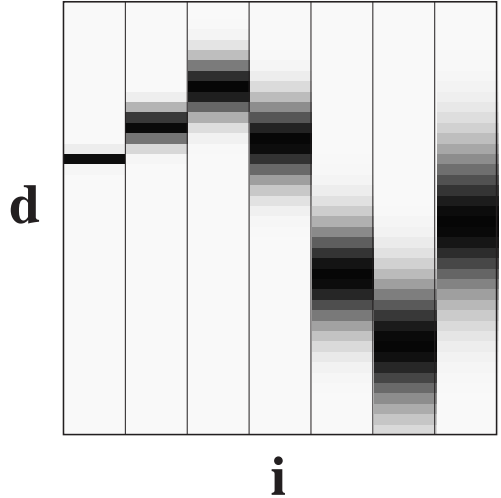}}
\caption{Dividing the space of independent parameters
in boxes, setting the
independent parameters to values that are certain to
belong to one of the boxes, 
performing the experiment for each of the possible ``boxes'' 
for the independent parameters, and, correspondingly,
measuring the values of the dependent parameters \ $ \bfd $ \ 
will produce states of information that are crudely 
represented in this figure. See text for an explanation.}
\label{fig: CondData}
\end{figure}

We have here assumed that the values of the independent
parameters are set randomly according to their null
information probability density. 
This directly leads to the ``Bayesian theory''
\ $ \theta(\bfi,\bfd) $ \ (this terminology being justified
in section~\ref{sec: UsingBayesTheory}).
A second option consists in defining physical 
correlations between parameters as a conditional probability density 
for the dependent parameters, given the independent parameters, 
\ $ \theta(\bfd|\bfi) $~,
but for the reasons explained elsewhere (\ref{note: CondProb}) 
the notion of conditional probability density, although a valid
mathematical definition, is not of direct use for handling
experimental results, unless enough care is taken.
Assume, for instance, that the space of independent parameters
is divided in boxes (multidimensional ``intervals'') and that the
independent parameters can be set to values that are certain to
belong to one of the boxes. 
Performing the experiment for each of the possible ``boxes'' 
for the independent parameters, and, correspondingly,
measuring the values of the dependent parameters \ $ \bfd $ \ 
will produce states of information that are crudely 
represented in figure~\ref{fig: CondData}. 
This collection of states of information correspond to the
conditional probability density \ $ \theta(\bfd|\bfi) $~.
The joint probability density in the \ $ (\bfi,\bfd) $ \ space
that carries this information without carrying any information
about the independent parameters (what we wish to call the
``Bayesian theory'') is then the product of the conditional
probability density \ $ \theta(\bfd|\bfi) $ \ by the null
information probability density for the independent paremeters,
say \ $ \mu_{\cal I}(\bfi) $~, i.e., the probability density
\begin{equation}
\theta(\bfi,\bfd) = \theta(\bfd|\bfi) \, \mu_{\cal I}(\bfi) \ .
\label{eq: TimesTheCond}
\end{equation}
To be more accurate, if, in each experiment, the only thing we know
about the independent parameters is the box where their value 
belongs, the measurement produces a probability density in the
\ $ (\bfi,\bfd) $ \ space, say \ $ \theta_i(\bfi,\bfd) $~, 
that equals the product of a probability density over 
\ $ \bfd $ \ (describing the result of the measurement
of the dependent parameters) times a probability density that equals
zero outside the box and equals the null information probability 
density inside the box. Applying the \OR\ operation to all these
probability densitues will also give the result of 
equation~\ref{eq: TimesTheCond}.

% % % % % % % % % % % % % % % % % % % % % % % % % % % % % % % % % % % % % %
% Begin Figure
% % % % % % % % % % % % % % % % % % % % % % % % % % % % % % % % % % % % % %

% % % % % % % % % % % % % % % % % % % % % % % % % % % % % % % % % % % % % %
% End Figure
% % % % % % % % % % % % % % % % % % % % % % % % % % % % % % % % % % % % % %

Interpreting the conditional probability density 
\ $ \theta(\bfd|\bfi) $ \
as simply
putting some ``error bars'' around some ``true functional
relationship'' \ $ \bfd = \bfd(\bfi) $~, that will always escape
to our knowledge, or assuming that the experimental knowledge 
\ $ \theta(\bfd|\bfi) $ \ 
represents is the ``real thing'', and that there is no necessity
of postulating the existence of a functional relationship, is a 
methaphysical question that will not change the manner of doing
physical inference. As explained in section \ref{sec: UsingBayesTheory},
inference will combine this ``theoretical knowledge'' 
represented by \ $ \theta(\bfi,\bfd) $ \ 
with further experiments using the \AND\ operation.

%\clearpage

% % % % % % % % % % % % % % % % % % % % % % % % % % % % % % % % % % % % % %
% Section
% % % % % % % % % % % % % % % % % % % % % % % % % % % % % % % % % % % % % %

\subsection{An example of Bayesian theory}

The discussion on the noninformative priors, in section~\ref{sec: NonInf}, 
was made without reference to a particular kind of object to be investigated.
Let us now turn to analyze the physics of the fall of objects at the surface 
of the Earth.

Assume we have a tube (with vacuo inside) of length \ $ L $ \ and we want
to analyze the time \ $ T $ \ it takes for a body 
to fall from the top to the bottom of the tube.
Experiments readily show that
\begin{equation}
L - \frac{1}{2} \, g \, T^2 \approx 0 \ ,
\end{equation}
where \ $ g $ \ is the acceleration of gravity at the given location,
but this ``law'' can not be exact for many reasons
i) residual air resistance;
ii) variation of gravity with height;
iii) relativistic effects;
iv) intrinsic (and so far unexplored) limitations of General Relativity;
etc.

We want to replace the line
\ $ L = \frac{1}{2} \, g \, T^2 $ \
by a probability density representing the actual knowledge that can 
be obtained from experiments. 
As explained in the previous section, the finite accuracy of any 
measurement will prevent the probability density from collapsing 
into a line ``without thickness''.

Let us face the actual problem of obtaining the probability density 
representing the theoretical/experimental knowledge on the physics of 
a falling body.
In the case where the length \ $ L $ \ is first 
set, and then the time \ $ T $ \ of the fall of the body measured
(this is, for instance, the way absolute gravimeters work, 
deducing, from the time \ $ T $~, the local value of 
the acceleration of gravity \ $ g $~; 
we will later face the alternative possibility), the 
experimenter should receive tubes of different lengths 
\ $ L_1 , L_2 \dots $ \ randomly generated according to the null 
information probability density for the length of an object,
i.e., with the probability density \ $ 1/L $~.

\begin{figure}[htbp]
\epsfxsize 10 cm
\centerline{\epsffile{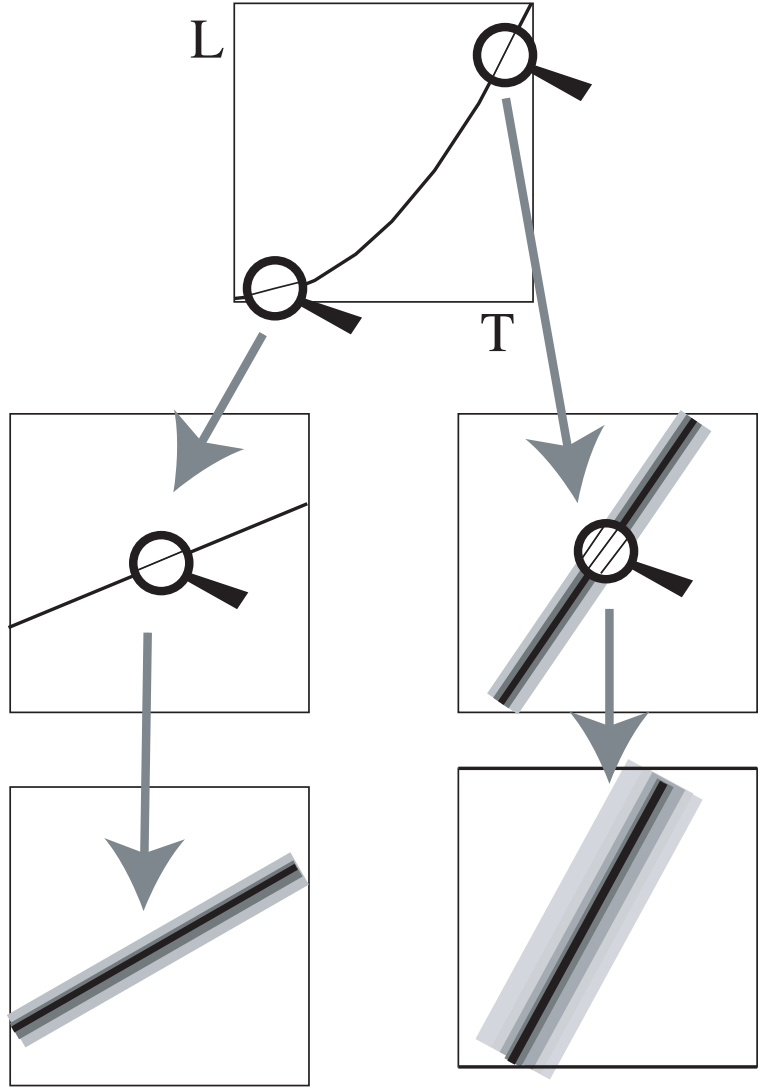}}
\caption{The free fall of an object inside a tube of length 
         \ $ L $ \ takes some time duration~$ T $~. 
         Experiments show that there is a good correlation 
         between \ $ L $ \ and \ $ T $~: with a good approximation, 
         \ $ L - \frac{1}{2} \, g \, T^2 \approx 0 $~. 
         An analitycal expression like 
         \ $ L = \frac{1}{2} \, g \, T^2 $ \ 
         can not be exact (any analytical theory 
         is just an approximation of reality).
         An examination of the real experiments 
         made to obtain the ``theory'' shows the 
         presence of uncompressible uncertainties. 
         Using the approach developed here, 
         the existing correlations between 
         \ $ L $ \ and \ $ T $ \ are represented 
         by a probability density which replaces 
         the classical notion of analytic theory. 
         If, at some scale, these correlations may 
         seem well described by an analytical expression 
         (here, in the top figure, by a line), succesive 
         magnifications (middle and bottom) end up by 
         showing the actual size of the ``theoretical 
         uncertainties''. In this example we have assumed 
         that measurements of lengths and of time durations 
         have constant relative errors (grossly exaggerated 
         in this schematic drawing). The thickness of this 
         theoretical distribution is of importance for: 
         i) solving, in a mathematical consistent manner 
         physical inference problems, and ii) accurately 
         computing uncertainties between physical parameters, 
         as, for instance, when predicting data values or 
         when solving inverse problems.}
\label{fig: Magnification} 
\end{figure}

When the first tube is provided to him, the experimenter should perform 
the falling experiment and, using the best possible equipment, measure 
as accurately as possible the length \ $ L $ \ of the tube given to
him and the time \ $ T $ \ it takes to the falling body to make the 
distance.
This would provide him with a probability density \ $ \theta_1(L,T) $ \ 
representing his knowledge of the realized value of the parameters.
There is no reason for the uncertainties on \ $ L $ \ and \ $ T $~,
as described by this probability density, to be independent.
When a second tube, with random length, is provided to him, he should 
perform again the experiment and obtain a second probability density 
\ $ \theta_2(L,T) $~.
As already explained, the ``Bayesian theory'' corresponding to 
these experiments is then the union (in the sense defined above) 
of all the states of information obtained 
in all the individual experiments, when their number tends to infinity:
\begin{equation}
\theta(L,T) = \sum_{i=1}^\infty \theta_i(L,T) \, .
\label{eq: AraTheo-1}
\end{equation}
Figure~\ref{fig: Magnification} schematizes the sort of 
probability density such a method would produce (\ref{note: LognoTheo}).

% % % % % % % % % % % % % % % % % % % % % % % % % % % % % % % % % % % % % %
% Begin Figure
% % % % % % % % % % % % % % % % % % % % % % % % % % % % % % % % % % % % % %

% % % % % % % % % % % % % % % % % % % % % % % % % % % % % % % % % % % % % %
% End Figure
% % % % % % % % % % % % % % % % % % % % % % % % % % % % % % % % % % % % % %

We have explored the case where the length \ $ L $ \ of the tube is 
first set and, then, the 
falling experiment is performed, measuring the time \ $ T $~.
The alternative is to fix the time duration \ $ T $ \ first and, then, 
to perform the the falling experiment, measuring the length \ $ L $ \ 
the falling body has traveled in that time.
The two sorts of experiments are not identical, as the type of 
measurements performed will be different and will lead to different 
uncertainties.
In this case, the experimenter is provided 
with time durations \ $ T_1 , T_2 \dots $ \ randomly selected according 
to the null information probability density 
for the period of a process, and obtains probability densities 
\ $ \vartheta_1(L,T) , \vartheta_2(L,T)  \dots $ \
representing the results of the measurements. The union of all these 
states of information
\begin{equation}
\vartheta(L,T) = \sum_{i=1}^\infty \vartheta_i(L,T) \, ,
\label{eq: AraTheo-2}
\end{equation}
would provide the ``Bayesian theory'' corresponding to that 
sort of experiment.

There is no reason for the two ``Bayesian theories'' thus obtained
to be identical, as they correspond to a different type of experiment.
We are then faced with the conclusion that the replacement of an 
analytical equation by a probability density will lead to 
probability densities attached to the precise experiment being performed.
In fact, this is not so different to what would have been obtained
when seeking for a functional relationship, as the ``best fitting
curve'' for the first kind of experiments may not be the ``best fitting''
one for the second kind of experiments.

The formation of a ``Bayesian theory'' here made by summing
small distributions (``histogramming'') can be understood in two ways.
First, we could perfectly well proceed in this way in practice,
performing systematic measurements of parameter correlations, 
using the best avalilable equipment.
Alternatively, we can understand the proposed method as a thought 
experiment helping to clarify what ``theoretical uncertainties'' can be.
These uncertainties can then be modeled using standard distributions 
(Gaussian, double exponential\dots) in such a way that usable
but still realistic probability distributions in the parameter 
space can be defined and used as ``Bayesian theories'', 
as in the example shown in note (\ref{note: LognoTheo}).

We will conclude this section with two remarks.
First, it is not possible to sample a probability distribution
that can not be normalized, as it is usually the case for the 
noninformative probabilities, like, for positive \ $ x $~, 
the \ $ 1/x $ \ distribution. Then, practical lower and upper bounds
have to be used.
Second, the number of experimental ``points'' that have
to be used in order to have a good practical approximation 
of a ``Bayesian theory'' depends on the accuracy of the 
measurements. Enough experiments have to be done so that 
the sum in equations~\ref{eq: AraTheo-1} 
and~\ref{eq: AraTheo-2} is smooth enough.
The sharper the experimental design, the more 
experiments we will need (and the mode detail we will have).

%\clearpage

% % % % % % % % % % % % % % % % % % % % % % % % % % % % % % % % % % % % % %
% Section
% % % % % % % % % % % % % % % % % % % % % % % % % % % % % % % % % % % % % %

\subsection{Using a Bayesian physical theory}
\label{sec: UsingBayesTheory}

Assume that enough experiments have been made, by skilled people,
using the best available equipment, following the guidelines
of the previous section, so that the ``Bayesian theory''
\ $ \theta(L,T) $ \ is available. Now a new tube 
is given to us, whose length has been randomly 
generated according to the null information probability density,
\ $ 1/L $~. 
We perform the falling experiment, perhaps with a more modest
equipment than that used to obtain the Bayesian theory, 
and measure the two parameters 
\ $ L $ \ and \ $ T $~, the result of the measurement being described
by the state of information \ $ \rho(L,T) $~.
How can we combine this information with the Bayesian theory, so
that we can ameliorate our knowledge on \ $ L $ \ and \ $ T $~?
We are exactly here in the situation where the notion of conditional
probability (in fact, our generalization of it) applies: we know that
we have a realization in the \ $ (L,T) $ \ space generated according
to the theoretical probability density \ $ \theta(L,T) $ \ 
and we have a state of information on this particular realization
that is described by the probability density \ $ \rho(L,T) $~.
The resulting state of information is then that obtained
by applying the \AND\ operation to these two states of information
(i.e., in the language defined above, by taking their intersection).
This gives
\begin{equation}
\sigma(L,T) = \frac{\theta(L,T) \, \rho(L,T)}{\mu(L,T)} \quad . 
\label{eq: InvBofill}
\end{equation}
In general, if \ $ \bfi $ \ is the independent parameter set and
\ $ \bfd $ \ the dependent set,
\begin{equation}
\sigma(\bfi,\bfd) = \frac{\theta(\bfi,\bfd) \, 
\rho(\bfi,\bfd)}{\mu(\bfi,\bfd)} \, .
\label{eq: TheGoodOne}
\end{equation}

If the information content concerning \ $ L $ \ contained in 
\ $ \rho(L,T) $ \ is very high (the length of the tube is well
known) while the information on \ $ T $ \ is low, then, 
\ $ \sigma(L,T) $ \ will essentially ameliorate our information
on \ $ T $~. This corresponds to the solution of a classical 
prediction problem in physics (how long it will take for a stone
to fall from the top of the tower of Pisa?).
Reciprocally, if the information content concerning 
\ $ T $ \ contained in 
\ $ \rho(L,T) $ \ is very high (the time of the fall is well
known) while the information on the length of the tube is low, then, 
\ $ \sigma(L,T) $ \ will essentially ameliorate our information
on \ $ L $~. Then, equation~\ref{eq: InvBofill} 
corresponds to the solution of an ``inverse problem'',
where ``data'' is used to infer the values of the parameters describing
some system.
This use of the notion of intersection of states of information to
solve inverse problems was advocated by 
Tarantola and Valette (\ref{note: TarantolaAndValette}) and
Tarantola (\ref{note: Tarantola}), who showed
that this method leads to results consistent with more particular
techniques (like least squares of least absolute values) when some
of the subtleties are ignored (theoretical uncertainties neglected, etc.).

We do not know of any alternative to our approach
that solves consistently nonlinear inverse problems.

%\clearpage

% % % % % % % % % % % % % % % % % % % % % % % % % % % % % % % % % % % % % %
% Section
% % % % % % % % % % % % % % % % % % % % % % % % % % % % % % % % % % % % % %

\section{Discussion and Conclusion}

Introducing Kolmogorov's definition of probability distributions
without introducing the two operations \OR\ and \AND, is like
introducing the real numbers without introducing the sum and the product:
we may compute, replacing clear mathematical objects by intuitive operations,
but we are lacking an important structure of the space.
The two operations we have introduced satisfy so obvious axioms that is 
difficult to imagine a simpler structure.

This structure may be used for many different inference problems,
but we have chosen here an illustration in the realm of physics.
We have replaced the notion of an analytical theory by the Bayesian 
notion of a probablity density representing all the experimentally 
obtained correlations between physical parameters, the space of 
independent parameters being visited randomly according to their 
null information probability density. Practically, some regions of 
the parameter space will not be accessible to investigation.
Accordingly, the ``result'' of the measurement will be the null 
information probability density for the corresponding parameters.
In other words, the ``error bars'' of a ``Bayesian theory'' may be large 
--- or even infinite --- for some regions of the parameter space. 
This is the typical domain where classical, analytical, theories 
extrapolate the equations that fit the observations made in a 
restricted region of the parameter space.
No such extrapolation is allowed with our approach.

Although we have only shown a simple example (the Galilean experiment),
the me\-thodology has a large domain of application.
As a further example, concerning tensor quantities,
we could examine the dependence between stress and strain for 
a given medium.
This would involve: i) mathematical definition of strain from
displacement; ii) operational definition of stress; and iii) 
analysis of the stress-strain correlation using the method 
described in this article.

Analytical theories, when extrapolating, predict results that may not 
correspond to observations, when they are made.
The theory is then ``falsified'' in the sense of Popper, and has to 
be corrected. A ``Bayesian theory'' can be indefinitely refined, as larger 
domains of the parameter space are accessible to experimentation, 
but never falsified.
The present work shows that pure empiricism (as opposed to the mathematical
rationalism of analytical theories) can be mathematically formalised.
This formalism is the only one known by the authors that handles
uncertainties consistently.

If physicists enjoy the game of extrapolation (as, for instance, 
when pushing Einstein's gravity theory to the conditions prevailing 
in a Big Bang model of the Universe), engineers advance by 
performing experiments as close as possible to the conditions 
that will prevail ``in the real thing''. 

Using the approach here proposed, the ``='' sign is only used
for mathematical definitions, as, for instance, when {\em defining\/}
a frequency from a period \ $ \nu = 1/T $~, or when using the mathematics
associated to probability calculus. But the ``='' sign is never used
to describle physical correlations, that are, by nature, only
approximate. These physical correlations are described by
probability distributions. Some may see the systematic use of the
``='' sign in mathematical physics as a misuse of mathematical 
concepts.

%\clearpage

% % % % % % % % % % % % % % % % % % % % % % % % % % % % % % % % % % % % % %
% References
% % % % % % % % % % % % % % % % % % % % % % % % % % % % % % % % % % % % % %

\section{References and Notes}

\def\bibref{\small\par\noindent\hangindent=7pt\rm\hskip 7 pt}

% % % % % % % % % % % % % % % % % % % % % % % % % % % % % % % % % % % % % %
% Note
% % % % % % % % % % % % % % % % % % % % % % % % % % % % % % % % % % % % % %

\begin{NoteAtEnd}
\bibref
Popper, K.R., 1934, Logik der forschung, Viena; English translation:
The logic of scientific discovery, Basic Books, New York, 1959.
\label{note: Popper}
\end{NoteAtEnd}

% % % % % % % % % % % % % % % % % % % % % % % % % % % % % % % % % % % % % %
% Note
% % % % % % % % % % % % % % % % % % % % % % % % % % % % % % % % % % % % % %

\begin{NoteAtEnd}
\bibref
Michelson, A.A. et E.W. Morley,
1887,
{\em Am. J. Sc.\/} (3),
{\bf 34},
333.
\label{note: MichelsonAndMorley}
\end{NoteAtEnd}

% % % % % % % % % % % % % % % % % % % % % % % % % % % % % % % % % % % % % %
% Note
% % % % % % % % % % % % % % % % % % % % % % % % % % % % % % % % % % % % % %

\begin{NoteAtEnd}
\bibref
Backus, G., 1970a, 
Inference from inadequate and inaccurate data: 
I,   Proc. Nat. Acad. Sci., 65, 1,   1--105;
II,  Proc. Nat. Acad. Sci., 65, 2, 281--287;
III, Proc. Nat. Acad. Sci., 67, 1, 282--289.
\label{note: Backus}
\end{NoteAtEnd}

% % % % % % % % % % % % % % % % % % % % % % % % % % % % % % % % % % % % % %
% Note
% % % % % % % % % % % % % % % % % % % % % % % % % % % % % % % % % % % % % %

\begin{NoteAtEnd}
\bibref
Tarantola, A., 1987, 
Inverse problem theory; methods for data fitting 
and model parameter estimation, 
Elsevier;
Tarantola, A.,
1990,
Probabilistic foundations of Inverse Theory,
in: {\em Geophysical Tomography\/},
Desaubies, Y., Tarantola, A., and Zinn-Justin, J., (eds.),
North Holland.
\label{note: Tarantola}
\end{NoteAtEnd}

% % % % % % % % % % % % % % % % % % % % % % % % % % % % % % % % % % % % % %
% Note
% % % % % % % % % % % % % % % % % % % % % % % % % % % % % % % % % % % % % %

\begin{NoteAtEnd}
\bibref
Tarantola, A., and Valette, B., 1982, 
Inverse Problems = Quest for Information, 
{\em J.~Geophys.\/}, 50, 159-170.
\label{note: TarantolaAndValette}
\end{NoteAtEnd}

% % % % % % % % % % % % % % % % % % % % % % % % % % % % % % % % % % % % % %
% Note
% % % % % % % % % % % % % % % % % % % % % % % % % % % % % % % % % % % % % %

\begin{NoteAtEnd}
\bibref
Mosegaard, K., and Tarantola, A.,
1995,
Monte Carlo sampling of solutions to inverse problems,
{\em J.~Geophys.\ Res.\/},
Vol.\ 100, No.\ B7, 12,431--12,447.
\label{note: MosegaardAndTarantola}
\end{NoteAtEnd}

% % % % % % % % % % % % % % % % % % % % % % % % % % % % % % % % % % % % % %
% Note
% % % % % % % % % % % % % % % % % % % % % % % % % % % % % % % % % % % % % %

\begin{NoteAtEnd}
\bibref
An expression \ $ d = d(m) $ \ connecting the data
\ $ d $ \ to the parameters \ $ m $ \ is typically used with 
the notion of conditional probability density 
(through the Bayes theorem) to make inferences. 
As discussed in note (\protect\ref{note: CondProb}),
conditional probability densities do not have the necessary invariant
properties when considering general (nonlinear) changes of variables.
\label{note: Unconsistency}
\end{NoteAtEnd}

% % % % % % % % % % % % % % % % % % % % % % % % % % % % % % % % % % % % % %
% Note
% % % % % % % % % % % % % % % % % % % % % % % % % % % % % % % % % % % % % %

\begin{NoteAtEnd}
\bibref
Kolmogorov, A.N., 
1933,
Grundbegriffe der Wahrcheinlichkeitsrechnung,
Springer, Ber\-lin;
Engl. trans.: Foundations of Probability,
New York, 1950.
\label{note: Kolmogorov}
\end{NoteAtEnd}

% % % % % % % % % % % % % % % % % % % % % % % % % % % % % % % % % % % % % %
% Note
% % % % % % % % % % % % % % % % % % % % % % % % % % % % % % % % % % % % % %

\begin{NoteAtEnd}
\bibref
Usually, the probability \ $ P(\calA) $ \ 
of a domain \ $ \calA $ \ is calculated via an expression like
\ $
P(\calA) = \int_\calA dM(\bfx) \, p(\bfx) \ ,
$ \ 
where \ $ M(\calA) $ \ is the {\em volume\/} (or {\em measure\/}) 
of \ $ \calA $~:
\ $
M(\calA) = \int_\calA dM(\bfx) \ .
$ \
The existence of the {\em volumetric probability\/} \ $ p(\bfx) $ \ 
is warranted by the Radon-Nicodym theorem if the
probability \ $ P $ \ is absolutely continous with respect to the measure
\ $ M $ \ (that is, if for any subdomain \ $ \calA $~, 
\ $ M(\calA) = 0 \Rightarrow P(\calA) = 0 $~).
Alternatively, one may write
\ $
M(\calA) = \int_\calA \underd \bfx \ \overmu(\bfx) 
$ \
and
\ $
P(\calA) = \int_\calA \underd \bfx \ \overp(\bfx) \ ,
$ \ 
where the {\em probability density\/} \ $ \overp(\bfx) $ \ 
is defined by
\ $ \overp(\bfx) = \overmu(\bfx) \, p(\bfx) $~.
The short notation \ $ \underd \bfx $ \ stands for \ $ dx_1 dx_2 \dots $~.
For instance, when considering a 3D Euclidean space with spherical 
coordinates,
\ $ \underd \bfx = dr \, d\theta \, d\phi $~,
\ $ \overmu(\bfx) = r^2 \, \sin\theta $~,
and
\ $ dM(\bfx) 
= \overmu(\bfx) \, \underd \bfx 
= r^2 \, \sin\theta \, dr \, d\theta \, d\phi$~.
In a change of variables, a probability density 
\ $ \overp(\bfx) $ \ 
is multiplied by the Jacobian of the transformation, 
while the associated volumetric probability \ $ p(\bfx) $ \ is invariant. 
The unfortunate gap existing between theoretical and practical 
presentations of probability theory induces frequent confusions between 
these two notions.
The choice of the reference measure \ $ M $ \ is obvious in 
geometrical spaces, as it is directly associated to the notion 
of volume. In more abstract spaces, like the spaces of physical 
parameters considered in this article, one has to introduce it 
explicitly. As explained elsewhere in the text, we interpret 
the probability density \ $ \overmu(\bfx) $ \ as representing 
the ``state of null information'' on the considered parameters 
(interpretation consistent with the absolute continuity 
postulated by the Radon-Nicodym theorem). 
In the main text we always consider probability densities, 
not volumetric probabilities, and, to simplify notations, 
the overlines and underlines of this note are not written.
\label{note: Radon-Nicodym}
\end{NoteAtEnd}

% % % % % % % % % % % % % % % % % % % % % % % % % % % % % % % % % % % % % %
% New definition of \bibref
% % % % % % % % % % % % % % % % % % % % % % % % % % % % % % % % % % % % % %

\def\bibref{\small\par\noindent\hangindent=7pt\rm}

% % % % % % % % % % % % % % % % % % % % % % % % % % % % % % % % % % % % % %
% Note
% % % % % % % % % % % % % % % % % % % % % % % % % % % % % % % % % % % % % %

\begin{NoteAtEnd}
\bibref
See, for instance, 
Kandel, A.,
1986,
Fuzzy mathematical techniques with applications,
Addison-Wesley.
\label{note: Kandel}
\end{NoteAtEnd}

% % % % % % % % % % % % % % % % % % % % % % % % % % % % % % % % % % % % % %
% Note
% % % % % % % % % % % % % % % % % % % % % % % % % % % % % % % % % % % % % %

\begin{NoteAtEnd}
\bibref
If \ $ f $ \ 
represents a probability density function in some coordinate system,
we will denote by \ $ f' $ \ the probability density in 
some transformed coordinates.
Under such a transformation, a probability density gets its values
multiplied by the Jacobian \ $ J $ \ of the transformation: 
\ $ f' = J \, f $~.
We have
$$
f' \vee g' =
(f \, J) + (g \, J) = (f + g) \, J =
(p \vee q)' \ ,
$$
which demonstrates the invariance of the \OR\ operation under a change
of variables.
If \ $ \mu $ \ represents the reference probability density
(neutral element for the \AND\ operation), we also have
$$
f' \wedge g' =
\frac{(f \, J) \, (g \, J)}{\mu \, J} = \frac{f \, g}{\mu} \, J =
(p \wedge q)' \ ,
$$
which demonstrates the invariance of the \AND\ operation under a change
of variables.
\label{note: DemoInvariance}
\end{NoteAtEnd}

% % % % % % % % % % % % % % % % % % % % % % % % % % % % % % % % % % % % % %
% Note
% % % % % % % % % % % % % % % % % % % % % % % % % % % % % % % % % % % % % %

\begin{NoteAtEnd}
\bibref
Once one has agreed on the form of the probability density
describing the state of null information, \ $ \mu(\bfx) $~,
Shannon's (\protect\ref{note: Shannon-2}) definition of 
information content of a probability density \ $ p(\bfx) $ \ 
has to be written
$$
I = \int_\calD \! d\bfx \, p(\bfx) \, \log\frac{p(\bfx)}{\mu(\bfx)} \ .
$$
Note that the ``definition'' 
\ $ I = \int_\calD \! d\bfx \, p(\bfx) \, \log p(\bfx) $ \ 
is not consistent, as it is not invariant under a change of variables.
\label{note: Shannon-1}
\end{NoteAtEnd}

% % % % % % % % % % % % % % % % % % % % % % % % % % % % % % % % % % % % % %
% Note
% % % % % % % % % % % % % % % % % % % % % % % % % % % % % % % % % % % % % %

\begin{NoteAtEnd}
\bibref
Shannon, C.E., 1948,
A mathematical theory of communication,
Bell System Tech. J., 27, 379--423.
\label{note: Shannon-2}
\end{NoteAtEnd}

% % % % % % % % % % % % % % % % % % % % % % % % % % % % % % % % % % % % % %
% Note
% % % % % % % % % % % % % % % % % % % % % % % % % % % % % % % % % % % % % %

\begin{NoteAtEnd}
\bibref
If \ $ \calA $ \ and \ $ \calB $ \ are two ``events''
(i.e., subsets of the space over which we consider a probability),
with respective probability \ $ P(\calA) $ \ and \ $ P(\calB) $~,
the conditional probability for the event \ $ \calA $ \ 
given the event \ $ B $ \ is defined by 
\ $ P(\calA | \calB) = P(\calA \cap \calB)/P(\calB) $~.
Consider, as an example, the Euclidean plane, with coordinates 
\ $ (x,y) $~. 
A probability distribution over the plane can be represented by a 
probability density \ $ p(x,y) $~. 
For finite \ $ \Delta x $ \ and \ $ \Delta y $~, 
one can consider the two events 
\ $
\calA = \left\{ \matrix{x_0 - \Delta x < x < x_0 + \Delta x \cr
                               -\infty < y < +\infty} \right\}
$ \ and \ $
\calB = \left\{ \matrix{       -\infty < x < +\infty \cr 
                        y_0 - \Delta y < y < y_0 + \Delta y } \right\}
$ \ 
representing respectively a ``vertical'' and an ``horizontal'' 
band of constant thicknesses \ $ \Delta x $ \ and \ $ \Delta y $ \ 
on the plane.
In normal circumstances, the ratio 
\ $ P(\calA \cap \calB) / P(\calB) $ \ 
has a finite limit when \ $ \Delta y \to 0 $~. 
For variable \ $ x $~, this defines a probability distribution 
over \ $ x $ \ 
whose density is named the ``conditional probability density over 
\ $ x $ \ given \ $ y = y_0 $~,'' and that is given by
$$
p(x|y=y_0) = \frac{p(x,y_0)}{\int_{-\infty}^{+\infty} \! dx \, p(x,y_0)} \ .
$$
It has to be realized that the probability density so defined 
depends on the fact that the limit is taken for a horizontal 
bar whose thickness tends to zero, this thickness being independent on 
\ $ x $~. Should we, for instance, have assumed a band around 
\ $ y = y_0 $ \ with a thickness being a function of \ $ x $~, 
we still could have defined a probability density, but it would 
not have been the same.
The problem with this appears when changes of variables are considered.
Changing for instance from the Cartesian coordinates \ $ (x,y) $ \ 
to some other system of coordinates \ $ (u,v) $ \ 
will change, according to the general rule, the (joint) probability 
density \ $ p(x,y) $ \ to 
\ $ q(u,v) = p(x,y) \left| \partial(x,y)/\partial(u,v) \right| $~. 
The line \ $ y = y_0 $ \ may become a line \ $ v = v(u) $~, 
but any sensible interpretation of an expression like
$$
q(u|v=v(u)) = \frac{q(u,v(u))}{\int_{u_{\rm max}}^{u_{\rm min}} 
\! du \, q(u,v(u))} \ .
$$
will consider a band {\em of constant thickness\/} \ $ \Delta v $ \ 
around the line \ $ v = v(u) $~. 
This band will not be (unless for linear changes of variable)
the transformed of the band considered when using the variables 
\ $ (x,y) $~.
This implies that any computation made using conditional probability 
densities in a given system of cordinates will not correspond to the 
use of conditional probability densities in other systems of coordinates.
Ignoring this fact leads to apparent paradoxes, as the so-called
Borel-Kolmogorov paradox, described in detail by 
Jaynes(\protect\ref{note: Jaynes}).
The approach we propose, where the notion of conditional probability 
is replaced by that of using the \AND\ operation on two probability 
distributions, is consistent with any change of variables, and will
not lead, even inadvertently, to any paradoxical result.
\label{note: CondProb}
\end{NoteAtEnd}

% % % % % % % % % % % % % % % % % % % % % % % % % % % % % % % % % % % % % %
% Note
% % % % % % % % % % % % % % % % % % % % % % % % % % % % % % % % % % % % % %

\begin{NoteAtEnd}
\bibref
Jaynes, E.T., 1995, Probability theory: the logic of science,
Internet (ftp: bayes.wustl.edu).
\label{note: Jaynes}
\end{NoteAtEnd}

% % % % % % % % % % % % % % % % % % % % % % % % % % % % % % % % % % % % % %
% Note
% % % % % % % % % % % % % % % % % % % % % % % % % % % % % % % % % % % % % %

\begin{NoteAtEnd}
\bibref
Cook, A., 1994, 
The observational foundations of physics,
Cambridge University Press.
\label{note: Cook}
\end{NoteAtEnd}

% % % % % % % % % % % % % % % % % % % % % % % % % % % % % % % % % % % % % %
% Note
% % % % % % % % % % % % % % % % % % % % % % % % % % % % % % % % % % % % % %

\begin{NoteAtEnd}
\bibref
Jeffreys, H., 1939,
Theory of probability,
Clarendon Press, Oxford.
\label{note: Jeffreys}
\end{NoteAtEnd}

% % % % % % % % % % % % % % % % % % % % % % % % % % % % % % % % % % % % % %
% Note
% % % % % % % % % % % % % % % % % % % % % % % % % % % % % % % % % % % % % %

\begin{NoteAtEnd}
\bibref
For instance, they can be taken equal to 
the standards of time duration and of frequency,
9~192~631 770 Hz and (1/9 192 631 770) s respectively.
\label{note: TheConstants}
\end{NoteAtEnd}

% % % % % % % % % % % % % % % % % % % % % % % % % % % % % % % % % % % % % %
% Note
% % % % % % % % % % % % % % % % % % % % % % % % % % % % % % % % % % % % % %

\begin{NoteAtEnd}
\bibref
The noninformative probability density for the position of a point
in an Euclidean space is easy to set in 
Cartesian coordinates:
\ $ p(x,y,z) = 1/V $~, where \ $ V $ \ is the volume of the
region into consideration. Changing coordinates, one can obtain the
form of the null information probability density in other coordinate 
systems.
For instance, in spherical coordinates, 
\ $ q(r,\theta,\varphi) = r^2 \, \sin \theta / V $~.
\label{note: Cartesian}
\end{NoteAtEnd}

% % % % % % % % % % % % % % % % % % % % % % % % % % % % % % % % % % % % % %
% Note
% % % % % % % % % % % % % % % % % % % % % % % % % % % % % % % % % % % % % %

\begin{NoteAtEnd}
\bibref
There is an amusing consequence to the fact that it is the
logarithm of the length (or the surface, or the volume)
of an object that is the natural (i.e., Cartesian) variable. 
The Times Atlas of the World 
(comprehensive edition, Times books, London, 1983) starts by listing
the surfaces of the states, territories, and principal islands 
of the world. The interesting fact is that the
{\em first digit\/} of the list is far from having an
uniform distribution in the range 1--9: the observed 
frequencies closely match the probability 
\ $ p(n) = \log_{10}\left((n+1)/n\right) $~,
(i.e., 30\%~of the occurrences are 1's, 
18\% are 2's,\dots,
and less than 5\% are 9's), that is the theoretical
distribution one should observe for a parameter whose
probability density is of the form \ $ 1/x $~. 
A list using the logarithm of the surface should
not present this effect, and all the digists 1--9
would have the same probability for appearing as 
first digit.
This effect explains the amusing fact first reported by 
Frank Benford in 1939: that the books containing tables
of logarithms (used, before the advent of digital computers, 
to make computations) have usually their first pages more 
damaged by use than their last pages\dots
\label{note: amusing}
\end{NoteAtEnd}

% % % % % % % % % % % % % % % % % % % % % % % % % % % % % % % % % % % % % %
% Note
% % % % % % % % % % % % % % % % % % % % % % % % % % % % % % % % % % % % % %

\begin{NoteAtEnd}
\bibref
Guide to the expression of uncertainty in measurement,
International Organization of Standardization (ISO),
Switzerland, 1993.
B.N. Taylor and C.E. Kuyatt, 1994, 
Guidelines for evaluating and expressing the uncertainty of NIST measurement results, NIST technical note 1297.
\label{note: ISO}
\end{NoteAtEnd}

% % % % % % % % % % % % % % % % % % % % % % % % % % % % % % % % % % % % % %
% Note
% % % % % % % % % % % % % % % % % % % % % % % % % % % % % % % % % % % % % %

\begin{NoteAtEnd}
\bibref
In priciple, all the parameters of the Universe are 
linked, and we could say that the only possible thing to do is to 
observe their time evolution. Even the free will of the experimenter
could be questioned.
We rather take here the empirical point of view that some parameters 
of the Universe can be discarded, some independent parameters set 
(i.e., an experiment defined), and that we can observe the effects 
of the experiment.
\label{note: FreeWill}
\end{NoteAtEnd}

% % % % % % % % % % % % % % % % % % % % % % % % % % % % % % % % % % % % % %
% Note
% % % % % % % % % % % % % % % % % % % % % % % % % % % % % % % % % % % % % %

\begin{NoteAtEnd}
\bibref
As a matter of fact, we have simply represented the probability
density
$$
\theta(L,T) = \frac{k}{L\,T} \, 
              \exp\left( - \frac{1}{2\,\sigma^2} 
              \left(\log\frac{L}{\frac{1}{2}\, g \, T^2} \right)^2\right) 
$$
for the value \ $ \sigma $ = 0.001~.
Its marginal probability densities are 
\ $ \theta_L(L) = \int_{T=0}^{T=\infty} dT \ \theta(L,T) = 1/L $ \
and
\ $ \theta_T(T) = \int_{L=0}^{L=\infty} dL \ \theta(L,T) = 1/T $~,
this meaning that the probability density \ $ \theta(L,T) $ \ 
carries no particular information on \ $ L $ \ and on \ $ T $~, 
but as this probability density takes significant values only when 
\ $ L \approx \frac{1}{2} \, g \, T^2 $~,
it carries all the information on the physical correlation between 
\ $ L $ \ and \ $ T $~.
\label{note: LognoTheo}
\end{NoteAtEnd}

% % % % % % % % % % % % % % % % % % % % % % % % % % % % % % % % % % % % % %
% Note
% % % % % % % % % % % % % % % % % % % % % % % % % % % % % % % % % % % % % %

\begin{NoteAtEnd}
\bibref
We thank Marc Yor for very helpful discussions concerning 
probability theory, and Dominique Bernardi for pointing 
to some important properties of real functions.
Enrique Zamora helped to understand grille's theory
from an engineer point of view.
B.N. Taylor and C.E. Kuyatt kindly sent us the very useful 
ISO's ``guide to the expression of uncertainty in measurement''.
This work has been supported in part by the French Minister of
National Education, the CNRS, and the Danish Natural Science Foundation.
\label{note: Thanks}
\end{NoteAtEnd}

\clearpage

\end{document}